# Comparative analysis of Debrecen sunspot catalogues

# Comparative analysis of Debrecen sunspot catalogues

L. Győri,[1]★ A. Ludmány[2]★ and T. Baranyi[2]★

[1]*Gyula Observing Station, DHO, Konkoly Observatory, Research Centre for Astronomy and Earth Sciences, Hungarian Academy of Sciences, Gyula, PO Box 93, H-5701, Hungary*
[2]*Debrecen Heliophysical Observatory (DHO), Konkoly Observatory, Research Centre for Astronomy and Earth Sciences, Hungarian Academy of Sciences, Debrecen, PO Box 30, H-4010, Hungary*

## ABSTRACT

Sunspot area data are important for studying solar activity and its long-term variations. At the Debrecen Heliophysical Observatory, we compiled three sunspot catalogues: the Debrecen Photoheliographic Data (DPD), the SDO/HMI Debrecen Data (HMIDD) and the SOHO/MDI Debrecen Data. For comparison, we also compiled an additional sunspot catalogue, the Greenwich Photoheliographic Data, from the digitized Royal Greenwich Observatory images for 1974–76. By comparing these catalogues when they overlap in time, we can investigate how various factors influence the measured area of sunspots, and, in addition, we can derive area cross-calibration factors for these catalogues. The main findings are as follows. Poorer seeing increases the individual corrected spot areas and decreases the number of small spots. Interestingly, the net result of these two effects for the total corrected spot area is zero. DPD daily total corrected sunspot areas are 5 per cent smaller than the HMIDD ones. Revised DPD daily total corrected umbra areas are 9 per cent smaller than those of HMIDD. The Greenwich photoheliographic areas are only a few per cent smaller than DPD areas. A 0.2° difference between the north directions of the DPD and MDI images is found. This value is nearly the same as was found (0.22°) by us in a previous paper comparing HMI and MDI images. The area measurement practice (spots smaller than 10 $mh$ were not directly measured but an area of 2 $mh$ was assigned to each) of the Solar Observing Optical Network cannot explain the large area deficit of the Solar Observing Optical Network.

**Key words:** methods: data analysis – catalogues.

## 1 INTRODUCTION

The area and position data of sunspots are widely used to study various aspects of solar activity, e.g. emergence, growth, and decay of spots; the connection between the structural development of a sunspot group and its flaring capability; solar irradiance variations; and periodicity in solar activity. For this reason, reliable long-term measurements are of great importance. At Debrecen Observatory, we have compiled several sunspot catalogues based on ground-based and space-borne solar images.

The Debrecen Photoheliographic Data (DPD) are mainly based on solar images taken in Gyula and Debrecen observatories (Győri, Baranyi & Ludmány 2011; Baranyi, Győri & Ludmány 2016). The gaps in the observations are filled in by images taken by observatories all over the word.

The SDO/HMI Debrecen Data (HMIDD) uses the magnetic and white-light images taken by the Helioseismic and Magnetic Imager (HMI; Schou et al. 2012) instrument on board the Solar Dynamics Observatory (SDO; Pesnell, Thompson & Chamberlin 2012).

The SOHO/MDI Debrecen data (SDD) are based on the magnetic and white-light images taken by the Solar and Heliospheric Observatory (SOHO; Scherrer et al. 1995) with the Michelson Doppler Imager (MDI) instrument.

DPD has common years with Greenwich Photoheliographic Results (GPR) on one side and with HMIDD on the other side. Moreover, DPD fully fills the gap between GPR and HMIDD. These two facts make it possible to calibrate the three catalogues to each other, and so to have a 140 yr-long collection of catalogues of sunspots. This makes it important to study the connections between DPD and GPR, and between DPD and HMIDD. The SDD and HMIDD catalogues have already been compared in a previous paper (Győri 2012).

Comparing the HMIDD and the DPD sunspot catalogues makes it possible to study how the seeing influences ground-based observations.

By comparing the heliographic coordinates of the sunspots in various catalogues, we can study the alignment differences between

★ E-mail: lajos.gyori49@gmail.com (LG); ludmany@tigris.unideb.hu (AL); baranyi@tigris.unideb.hu (TB)

the images of these catalogues, and, through this, the accuracy of the sunspot positions.

## 2 METHODS

For clarity, we define some terms used in the paper. Sunspot area is the area of the whole spot (umbra plus penumbra) or the area of a pore (a small sunspot without a penumbra, which is lighter than an umbra). The total sunspot area is the area of the sunspots summed over the whole solar disc. The umbra area is the area of the umbra of the sunspot, and pores are excluded. The total umbra area is the area of the umbrae summed over the whole solar disc. Projected areas are measured in millionths of the solar disc (*md*) and corrected areas in millionths of the solar hemisphere (*mh*).

The relationship between the areas of a solar feature in two catalogues ($C_1$ and $C_2$) is modelled by linear regression in the form $A_2 = aA_1$, where $A_1$ and $A_2$ are the areas in $C_1$ and in $C_2$, respectively.

We provide three types of plots to show details of the regression: the scatter plot of the regression, the moving root mean square residual (MRMR) and the moving relative root mean square residual (MRRMR).

MRMR measures how the root mean square residual (RMR) relative to the regression line depends on the area. For a given area $A$, the MRMR is determined from the $n$-nearest areas to $A$. Similarly, MRRMR measures how the relative root mean square residual (RRMR) relative to the regression line depends on the area. For a given area $A$, the MRRMR is determined from the $n$-nearest areas to $A$. We choose the value 20 for $n$.

To compile the various sunspot catalogues compared in this paper, we used the software package SAM (Sunspot Automatic Measurement). SAM is a set of cooperative computer programs that embraces every aspect of compiling a sunspot catalogue, from (1) setting up the necessary data base for the observational data of the solar images and the telescopes and (2) automatically detecting sunspots (umbra and penumbra) on solar images and determining their heliographic coordinates and area through (3) to making the catalogue ready for publishing (Győri 1998).

To identify sunspots on solar images, we used the method published by Győri (2015). A solar image may contain not only features belonging to the Sun but also various artefacts too, such as emulsion deficiency, emulsion abrasion, scratches, dust grains, clouds, inscriptions, patches, orientation markers, and much more. The properties of some of these artefacts are like sunspots (especially when the image quality is not too good), so it is hard to filter them out automatically, but a trained human eye/brain can do this in many cases (not always). Even when the average seeing of the image is good, there are regions with poorer seeing and small sunspots in these regions cannot be surely perceived. However, if, in a close-in-time image, these spots are situated in regions with good seeing, then they will show up clearly and could provide conformation. Near the solar limb, especially when the image quality is not too good, the penumbra and the umbra borders are not too definitive. In this case, the human intervention could increase the accuracy of finding these borders. To cope with these situations and with others not mentioned here, the automated sunspot catalogue created by SAM can be humanly revised by using special software, SAMm (Sunspot Automatic Measurement modification), developed for this purpose.

SAMm shows the penumbral and umbral borders of the spots, and the sunspot gravity centres superimposed on the solar image as they were found by SAM. If the reviser thinks so or another close-in-time image supports it, these borders can be modified. Moreover, pores, spots and umbrae can be deleted or added. The revision requires a lot of human resources and a lot of time. However, by involving additional solar images in the image processing, the accuracy of the solar catalogue can be increased.

The SDO/MDI and HMIDD catalogues are issued in an automated version. DPD catalogues come out in two versions: a preliminary (automated) one and a revised one.

## 3 EFFECT OF IMAGE QUALITY ON SOLAR FEATURE AREA

The factors that influence the result of the area measurement of the solar features can be divided into two main parts: the measuring method and the quality of the solar image. The solar image quality is determined by several factors: the optical resolution of the telescope (Rayleigh limit), the image scale, the properties of the image acquisition media (film or CCD) and the seeing (for ground-based observations).

For a ground-based observation, it is the seeing that mostly influences the image quality. For photoheliograms, beside the seeing, the gamma of the film, the exposure (underexposed, well or overexposed), and the deterioration of the film developer can have a major effect on image quality.

Image quality can influence the measured area of a solar feature (umbra, spot or pore) in three interplaying ways:

(i) *Border inaccuracy*. A solar feature's border has a ragged structure as we know from high-scale and high-quality solar images. How this ragged structure manifests in a solar image depends on the quality of the image and, therefore, the measured area may also depend on this. Similarly, poorer image quality adversely influences the overall definition of borders of the solar features, and so decreases the accuracy of the area measurement.

(ii) *Structural adequacy*. Image quality may also affect the structure of the solar features, and through this it can influence the measured area too. For example, long thin fissures intruding a feature do not appear in a poorer quality image, and so its area is included in the area of the feature. Similarly, nearby standalone features separated by a long narrow gap may merge in a poorer quality image and the area of the gap is included in the area of the merged features.

(iii) *Feature perceivability*. Poorer image quality may cause small low-contrast features (pores, small umbrae and dying parts of a larger penumbra) to dissolve into their environment, and so they cannot be perceived.

Thoroughly comparing the top and the bottom rows of Figs 1 and 2, we can observe the above described effects in action. We can also see from these figures that, although, the Gyula images have better image scale than those of SDO/HMI ones, their overall image qualities are perceivably lower even in good seeing conditions, and much lower in very bad seeing conditions. Notice that the SDO/HMI and Gyula images in Figs 1 and 2 are practically taken at the same time (9 s time difference for Fig. 1 and 12 s time difference for Fig. 2), so differences in the sunspots are caused only by the differences in the image quality.

## 4 HMI AND AUTOMATED DPD COMPARISON

The most adverse factor for a ground-based solar observation is the seeing (the random component of the astronomical refraction, caused by fluctuations in the refractive index along the light path). Comparing sunspot areas measured in almost co-temporal (below

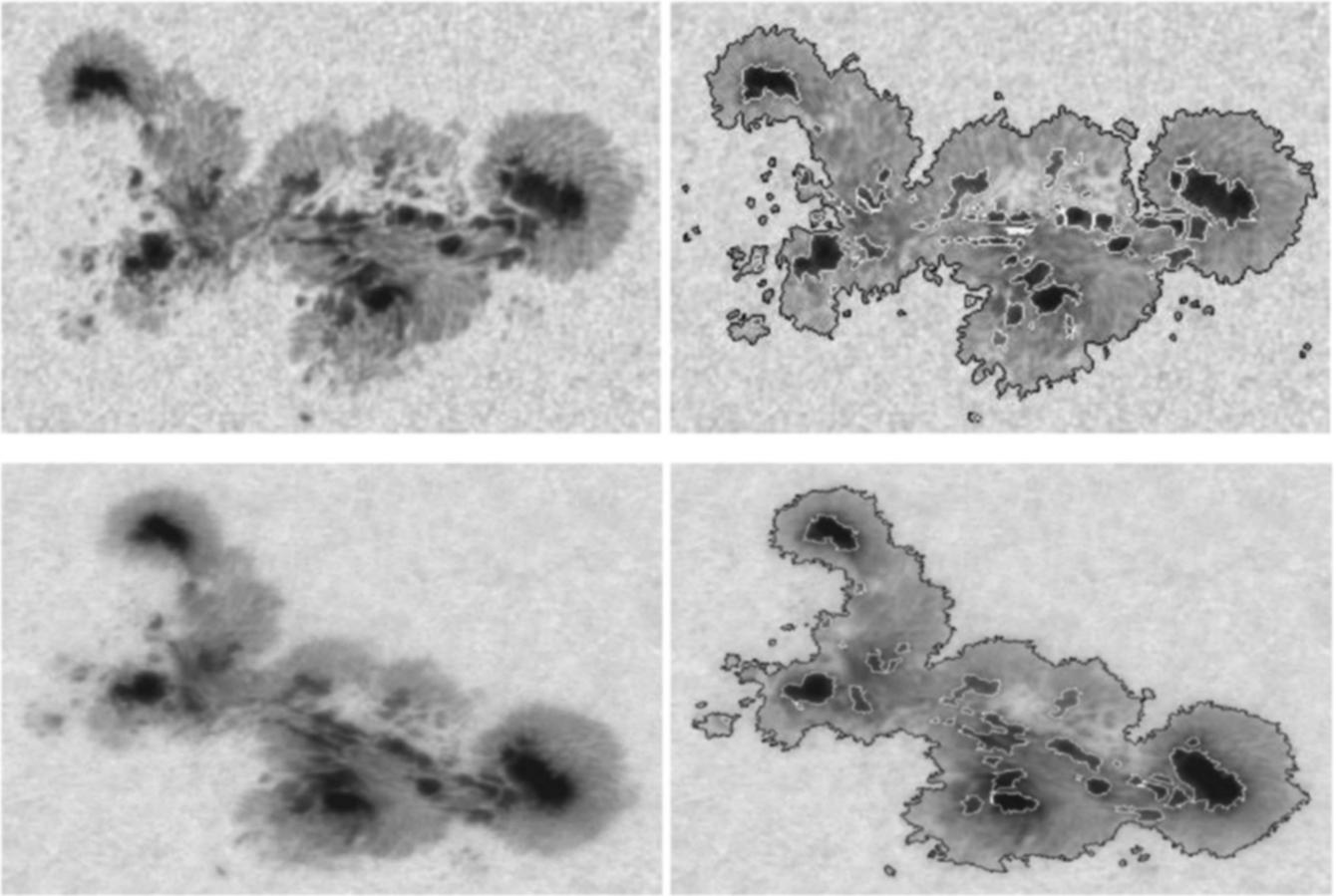

**Figure 1.** Top left, image of NOAA AR 11429 taken on 2012 March 7 at 11:03:29 UT by SDO/HMI. Top right, the same as top left, but with the boundary pixels (black penumbra, white umbra) superimposed on the image. Bottom row, the same sunspot group as in the top row, taken 9 s later by Gyula Observatory in good seeing condition. Note that the photospheric areas inside a penumbra border are found by SAM and their areas are not included in the area of the spot. For how the penumbra and umbra boundaries are determined, see Győri (1998).

1 min apart) HMI and DPD images provides an excellent possibility for studying how the seeing affects the measured area of the sunspots. We will compare HMI and DPD spot areas for these seeing qualities (SQ): very bad (VB), bad (B), medium (M), good (G) and very good (VG). The determination of the seeing quality of an image is automatic, and based on how well the solar granulation can be perceived in the solar image. We intend to publish the method in the near future.

We also want to examine how the different perceivabilities of the small spots in the HMI and DPD images affect the area discrepancy between DPD and HMI. Moreover, we want to examine how the decreasing intensity contrast of the photosphere and the decreasing resolution (relative to the solar surface) of the solar image as one goes from the centre to the limb (the foreshortening effect) influence the area discrepancy between DPD and HMI.

Thus, we determine the regression between the HMI and the DPD sunspot areas by using the following spot selections for the spot areas ($A$) and for the heliocentric angle ($\gamma$):

(i) $A > 0$ and $\gamma < 90°$. This actually means no selection, either for spot area or for heliocentric angle of the spot.

(ii) $A > 5\,mh$ and $\gamma < 90°$. Only the sunspots with area larger than $5\,mh$ are taken into account, and there is no selection for the heliocentric angle of the spot.

(iii) $A > 0\,mh$ and $\gamma < 60°$. There is no selection for the spot area, but only the spots with heliocentric angle smaller than $60°$ are taken into account.

(iv) $A > 5\,mh$ and $\gamma < 60°$. Only the sunspots with area larger than $5\,mh$ and with heliocentric angle smaller than $60°$ are taken into account.

For the investigations outlined in the above paragraphs, we compiled automatically two special full-disc sunspot catalogues (no separation of the spots into sunspot groups): the DPD and the HMI. For the DPD catalogue, all 2087 images taken in 2012 in Gyula were used. The HMI catalogue was based on the HMI quasi-continuum and the magnetic images that were near co-temporal with the DPD ones. The pores in a HMI image were automatically checked against the image taken 10 min later. If a pore existed in this image too, then it was retained, if not, then it was discarded.

### 4.1 Feature number comparison

When we compare feature numbers, we are mostly sensitive to feature perceivability.

Table 1 shows the ratio of the various feature numbers for the HMI images to those of the DPD images for various seeing qualities. The various feature number ratios depend on the seeing. Poor seeing increases the ratio, i.e. decreases the number of features observed in the DPD images. Even during good seeing conditions, significantly

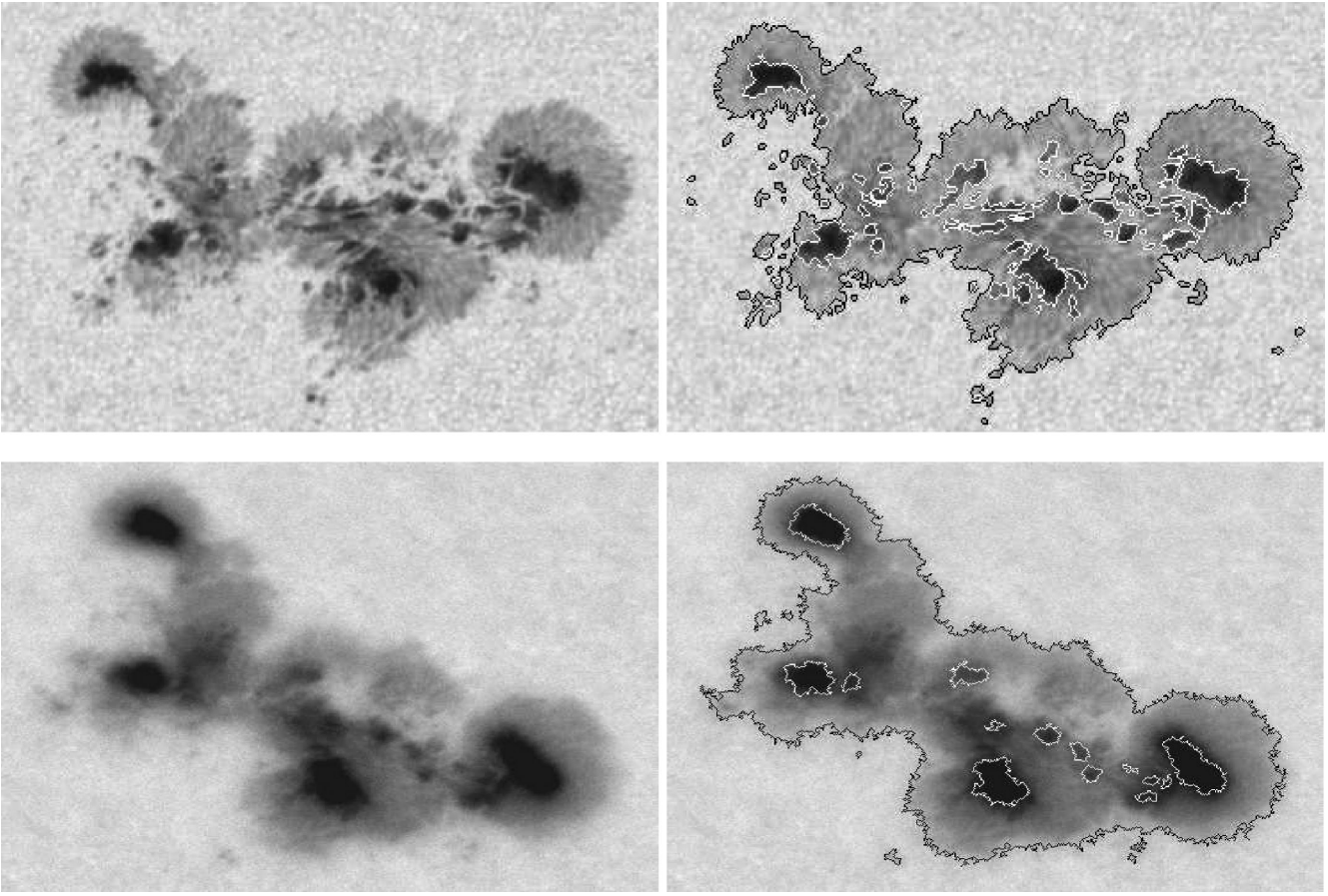

**Figure 2.** Top left, image of NOAA AR 11429 taken on 2012 March 7 at 13:52:14 UT by SDO/HMI. Top right, the same as the top left, but with the boundary pixels (black penumbra, white umbra) superimposed on the image. Bottom row, the same sunspot group as in the top row, taken 12 s later by Gyula Observatory in very bad seeing condition. Note that the photospheric areas inside a penumbra border are found by SAM and their areas are not included in the area of the spot. For how the penumbra and umbra boundaries are determined, see Győri (1998).

**Table 1.** Ratio of the various feature numbers for the HMI images to those of the DPD images for various seeing qualities.

| SQ | Pore ratio | Penumbra ratio | Umbra ratio | Spot ratio |
|---|---|---|---|---|
| VG | 1.77 | 1.83 | 1.70 | 1.78 |
| G  | 1.91 | 1.97 | 1.85 | 1.93 |
| M  | 2.19 | 2.17 | 1.99 | 2.19 |
| B  | 2.61 | 2.31 | 2.08 | 2.56 |
| VB | 4.47 | 2.67 | 2.23 | 3.84 |

**Table 2.** Number of the various features in the images used to compile Table 1 for the HMI and for the DPD images.

|  | Pore number | Penumbra number | Umbra number | Spot number |
|---|---|---|---|---|
| HMI | 152 422 | 46 071 | 86 695 | 198 493 |
| DPD | 60 690  | 20 346 | 42 591 | 81 036  |

more features can be observed in the HMI images than in the DPD ones.

To give an idea of the sample size used to compile Table 1, Table 2 provides the number of the various solar features in the HMI and in the DPD images.

### 4.2 Individual corrected area comparison

If we compare individual features, we are not sensitive to the *feature perceivability* effects. In this case, it is only the *border inaccuracy* and the *structural adequacy* that matter.

#### 4.2.1 Individual corrected spot area comparison

50 309 individual sunspots were identified in the nearly co-temporal HMI and DPD image pairs. The regression between HMI and DPD individual corrected spot areas was determined for various seeing qualities of the DPD images. The regression was also carried out with and without the selection that the heliocentric angle of the spot is smaller than $60°$. The independent variable is the HMI area. Table 3 summarizes the results.

We can see from Table 3 that the poorer the seeing, the larger (relative to the HMI) the DPD individual corrected spot area. The individual corrected spot areas in the two catalogues are practically the same for VG, G and M seeing qualities, but for very bad seeing, the DPD individual corrected spot areas are larger by 4 per cent ($\gamma < 90°$) and by 5 per cent ($\gamma < 60°$) than the HMI ones. Moreover, we can observe that the foreshortening effect slightly moderates the increase of the DPD individual corrected spot areas with the poorer seeing (compare columns 2 and 3), because the slopes are slightly smaller for $\gamma < 90°$ than for $\gamma < 60°$.

**Table 3.** Summary of the results of the comparison of the individual corrected sunspot areas measured in HMI and in DPD images. The values in the table are the slopes of the regression lines at the proper seeing and spot sections. The independent variable is the HMI area. See the text for more details.

| SQ | $\gamma < 90°$ | $\gamma < 60°$ |
|---|---|---|
| VG | 0.994 | 1.006 |
| G  | 1.001 | 1.013 |
| M  | 1.004 | 1.012 |
| B  | 1.015 | 1.022 |
| VB | 1.041 | 1.048 |

**Table 4.** Summary of the results of the comparison of the individual corrected umbrae areas measured in HMI and in DPD images. The values in the table are the slopes of the regression lines at the proper seeing and spot selection. The independent variable is the HMI area. See the text for more details.

| SQ | $\gamma < 90°$ | $\gamma < 60°$ |
|---|---|---|
| VG | 0.929 | 0.931 |
| G  | 0.934 | 0.931 |
| M  | 0.925 | 0.919 |
| B  | 0.917 | 0.910 |
| VB | 0.935 | 0.927 |

**Table 5.** Summary of the results of the comparison of the total corrected sunspot areas measured in HMI and in DPD images. The values in the table are the slopes of the regression lines at the proper seeing and spot selection. The independent variable is the HMI area. See the text for more details.

| SQ | $A > 0\,mh$ $\gamma < 90°$ | $A > 5\,mh$ $\gamma < 90°$ | $A > 0\,mh$ $\gamma < 60°$ | $A > 5\,mh$ $\gamma < 60°$ |
|---|---|---|---|---|
| VG | 0.893 | 0.927 | 0.949 | 0.993 |
| G  | 0.899 | 0.932 | 0.953 | 1.001 |
| M  | 0.895 | 0.940 | 0.946 | 1.005 |
| B  | 0.897 | 0.948 | 0.955 | 1.023 |
| VB | 0.900 | 0.967 | 0.961 | 1.045 |

#### 4.2.2 Individual corrected umbra area comparison

23 694 individual umbrae were identified in the nearly co-temporal HMI and DPD image pairs. We applied the same selection criteria for individual umbrae as in the section above for individual spots.

Table 4 exhibits the slopes of the regression lines for various seeing qualities and $\gamma$ selections. The DPD individual corrected umbrae areas are about 7 per cent smaller than the HMI ones. Moreover, no definite seeing and $\gamma$ dependence (foreshortening effect) was found.

### 4.3 Total corrected area comparison

Total corrected sunspot and umbra areas were determined for 2087 nearly co-temporal HMI and DPD image pairs. The total sunspot or total umbra area derived from a solar image is influenced by all three ways that image quality can affect the area measurement (border inaccuracy, structural adequacy and feature perceivability).

#### 4.3.1 Total corrected spot area comparison

Table 5 summarizes the results of the comparison of the total corrected sunspot areas measured in HMI and in DPD images. The comparison was carried out with various spot selections applied to the sunspots.

**Table 6.** Summary of the results of the comparison of the total corrected umbra areas measured in HMI and in DPD images. The values in the table are the slopes of the regression lines at the proper seeing and spot selection. The independent variable is the HMI area. See the text for more details.

| SQ | $A > 0\,mh$ $\gamma < 90°$ | $A > 5\,mh$ $\gamma < 90°$ | $A > 0\,mh$ $\gamma < 60°$ | $A > 5\,mh$ $\gamma < 60°$ |
|---|---|---|---|---|
| VG | 0.790 | 0.854 | 0.832 | 0.901 |
| G  | 0.756 | 0.831 | 0.812 | 0.905 |
| M  | 0.741 | 0.827 | 0.786 | 0.886 |
| B  | 0.737 | 0.845 | 0.771 | 0.898 |
| VB | 0.730 | 0.861 | 0.740 | 0.898 |

The fifth column ($A > 5\,mh$ and $\gamma < 60°$) shows the case when we are not sensitive to two effects: the perceivability effect by leaving out small spots from both catalogues and the effect generated near the limb (foreshortening). For VG, G and M seeing qualities, there is practically no difference between the total corrected sunspot areas in the HMI and the DPD catalogues. We can also observe that the poorer seeing increases the area difference by making the DPD area larger, e.g. when the seeing is very bad the DPD total corrected sunspot areas are 4 per cent larger than the HMI ones.

The fourth column ($A > 0\,mh$ and $\gamma < 60°$) shows how excluding the foreshortening effect influences the area difference between the two catalogues. There is an about 5 per cent DPD area deficit. No definite seeing influence was found. This can be explained by the counterbalance between two types of seeing effects. First, the poorer seeing decreases the number of detectable small spots and so the total spot area. Secondly, the poorer seeing increases the individual spot area (Section 4.2.1).

In Table 3, the slope is larger than 1 because individual spot areas are larger with DPD. However, since HMI finds more spots, this last effect dominates here in column 4 (slope lower than 1). It does not change in column 5 because of the $A$ threshold (the additional spots with HMI are mostly small).

The third column ($A > 5\,mh$ and $\gamma < 90°$) shows what happens if spots smaller than $5\,mh$ are left out from both catalogues. In this case, we experience a DPD area deficit that depends on the seeing. This seeing dependency can be attributed to the seeing dependency of the individual spots (Section 4.2.1).

From the second column ($A > 0$ and $\gamma < 90°$) of Table 5, we see that, when we do not apply any selection at all, there is a 10 per cent DPD area deficit. There is no seeing dependency. This lack of seeing dependency can be explained similarly to the one found in the fourth column of Table 5, namely by compensating for the fewer small spots with the larger individual corrected spot areas.

#### 4.3.2 Total corrected umbra area comparison

When we compare total corrected umbra areas between DPD and HMI, we proceed the same way as for comparing total corrected spot area. Table 6 summarizes the results of the comparison.

The values in the table are the slopes of the regression lines at the proper seeing and selection. The independent variable of the regression is the HMI total corrected umbra area.

The fifth column ($A > 5\,mh$ and $\gamma < 60°$) shows that by leaving out the small umbrae and the umbrae in the outer part of the solar disc, the area deficit is only 10 per cent and no seeing dependence

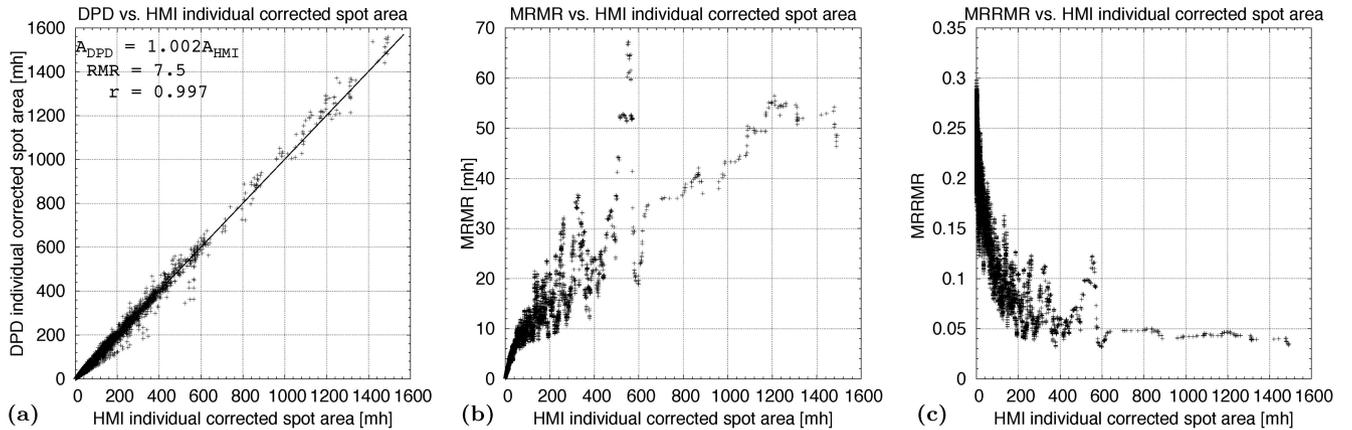

**Figure 3.** (a) Individual corrected spot areas in millionths of the solar hemisphere (*mh*) for 2012, for DPD versus HMI (pluses) and the regression line fitted to the data (solid line). (b) MRMR versus HMI individual corrected spot areas. (c) MRRMR versus HMI individual corrected spot areas. DPD images for the very good, the good and the medium seeing quality are used to create these figures, and no selections are applied to the spot area and to the heliocentric angle ($A > 0\,mh$ and $\gamma < 90°$).

is present in the data (similarly as for the individual umbrae in Table 4).

The fourth column ($A > 0\,mh$ and $\gamma < 60°$) shows the case when the umbrae in the outer part of the solar disc are excluded for both catalogues. The area deficit decreases compared to the second column ($A > 0$ and $\gamma < 90°$), but depends on the seeing in a similar way. The larger DPD umbra area deficit in the second column relative to those in the fourth column can be explained by the common effect of the seeing and the lower intensity contrast near the limb: the poor seeing even further decreases the intensity contrast near the limb, and, therefore, fewer small umbrae can be detected.

The third column ($A > 5\,mh$ and $\gamma < 90°$) shows that by leaving out the small umbrae from both of the catalogues the area deficit significantly decreased. In this case, no definite dependence of the area deficit on the seeing was found. This indicates that the seeing influences the total corrected umbra area through the perceivability of the small umbrae.

The second column ($A > 0$ and $\gamma < 90°$) of Table 6 shows the case when no selection is applied to the umbrae. The DPD total umbra areas are smaller (DPD total umbra area deficit) compared to the HMI ones. This area deficit depends on the seeing. The poorer the seeing, the larger the area deficit, e.g. for very good seeing the DPD total umbra area is smaller by 21 per cent than the HMI one, and, for very bad seeing, this value amounts to 27 per cent.

We conclude that the influence of seeing on umbra areas is twofold: it decreases the detected number of small umbrae (and so the total umbra area) and, interplaying with the foreshortening, even further decreases the detected number of umbrae in the outer part of the solar disc.

### 4.4 Details of the area regression between HMI and automated DPD

There are 60 kinds of comparison between the HMI and the DPD areas in the previous parts of this section. This number is too large to include the figures visualizing the finer details of the regression between the HMI and the DPD areas in this paper. However, these figures would be important for assessing the accuracy of the regression. Therefore, we provide these figures for images falling in one of the very good, the good and the medium seeing qualities. We choose these seeing categories because the images used for compiling the revised DPD catalogue mainly fall into these categories as we will see in Section 5. Even in this case, we provide the figures only for the case when there is no selection on the area and on the heliocentric angle of the spots ($A > 0\,mh$ and $\gamma < 90°$).

Fig. 3 shows the details of the regression between individual corrected spot areas from DPD images with very good, good and medium seeing qualities and from nearly co-temporal HMI images. Fig. 3(a) shows the scatter plot and the regression line fitted to the data. Fig. 3(b) shows the MRMR versus HMI individual corrected spot areas. Fig. 3(c) shows the MRRMR versus HMI individual corrected spot areas. Fig. 3(b) shows that MRMR increases, with more or less fluctuation, with the area. Fig. 3(c) exhibits that, for small areas, MRRMR rapidly decreases with the area. Then, after a slower decrease, it becomes nearly constant (0.05) for areas above 200–400 *mh*.

Furthermore, Fig. 3(a) shows a few large residuals at about 560 *mh*. This anomaly shows up as a significant regional maximum in Fig. 3(b) and in Fig. 3(c). This anomaly is caused by a large spot just rotating on to the east limb. By visually examining the spot border contours found by SAM in the corresponding DPD and HMI images, we found good agreement with the spot border for both of the DPD and HMI images. The heliocentric angle of the spot is 82.7°. At this position, $1/\cos\gamma = 7.84$. So even a small inaccuracy in the spot border results in a much larger inaccuracy in spot area when transformed on the solar sphere.

Fig. 4 shows the same plots for individual corrected umbra areas. The tendencies observed for MRMR and for MRRMR in Fig. 4(b) and in Fig. 4(c) are like those observed in Fig. 3(b) and in Fig. 3(c) (although the flat part is not as obvious).

Fig. 5 shows the same plots for total corrected spot areas. Fig. 5(b) shows that MRMR increases, with more or less fluctuation, with the area. Fig. 5(c) exhibits that MRRMR decreases (with large fluctuations) with the area.

Fig. 6 shows the same plots for total corrected umbra areas. The tendencies observed for MRMR and for MRRMR in Fig. 6(b) and in Fig. 6(c) are like those observed in Fig. 5(b) and in Fig. 5(c).

Table 7 summarizes the parameters of the various area regressions detailed by the corresponding figures of this section. If we compare the RMR of regression with the MRMR shown by the corresponding

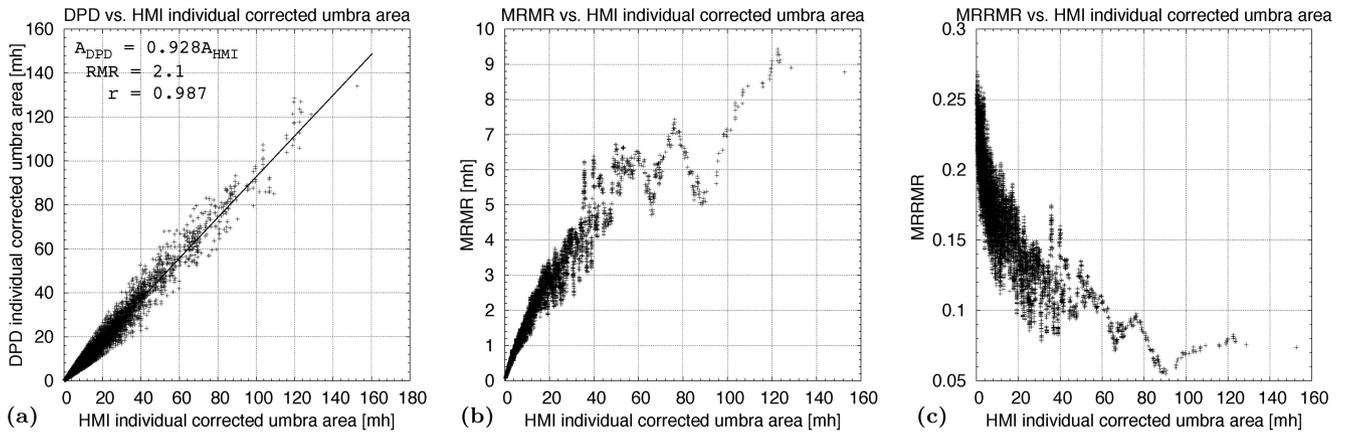

**Figure 4.** (a) Individual corrected umbra areas in millionths of the solar hemisphere (*mh*) for 2012, for DPD versus HMI (pluses) and the regression line fitted to the data (solid line). (b) MRMR versus HMI individual corrected umbra areas. (c) MRRMR versus HMI individual corrected umbra areas. DPD images for the very good, the good and the medium seeing quality are used to create these figures and no selections are applied to the umbra area and to the heliocentric angle ($A > 0\,mh$ and $\gamma < 90°$).

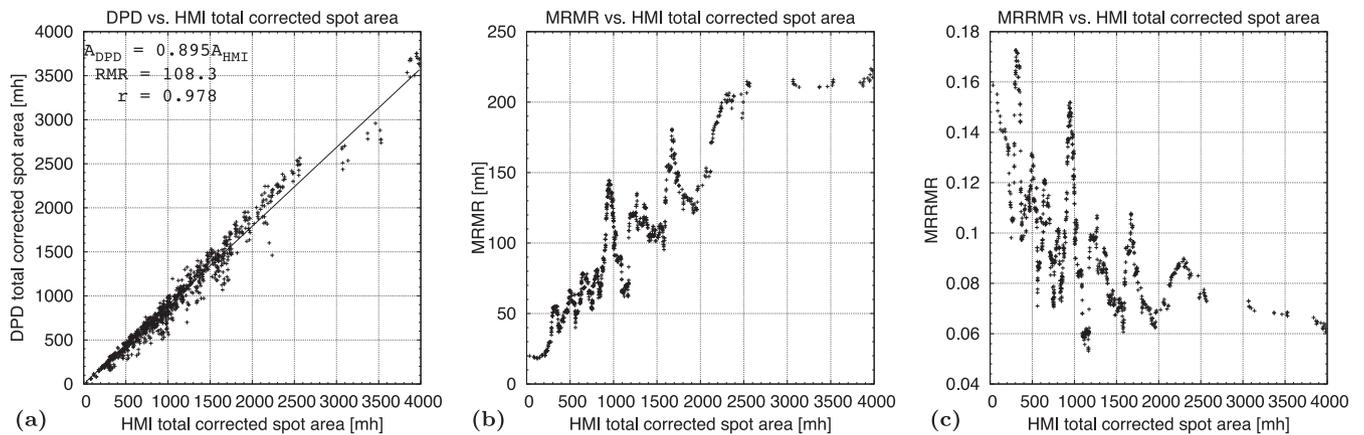

**Figure 5.** (a) Total corrected spot areas in millionths of the solar hemisphere (*mh*) for 2012, for DPD versus HMI (pluses) and the regression line fitted to the data (solid line). (b) MRMR versus HMI total corrected spot areas. (c) MRRMR versus HMI total corrected spot areas. DPD images for the very good, the good and the medium seeing quality are used to create these figures and no selections are applied to the spot area and to the heliocentric angle ($A > 0\,mh$ and $\gamma < 90°$).

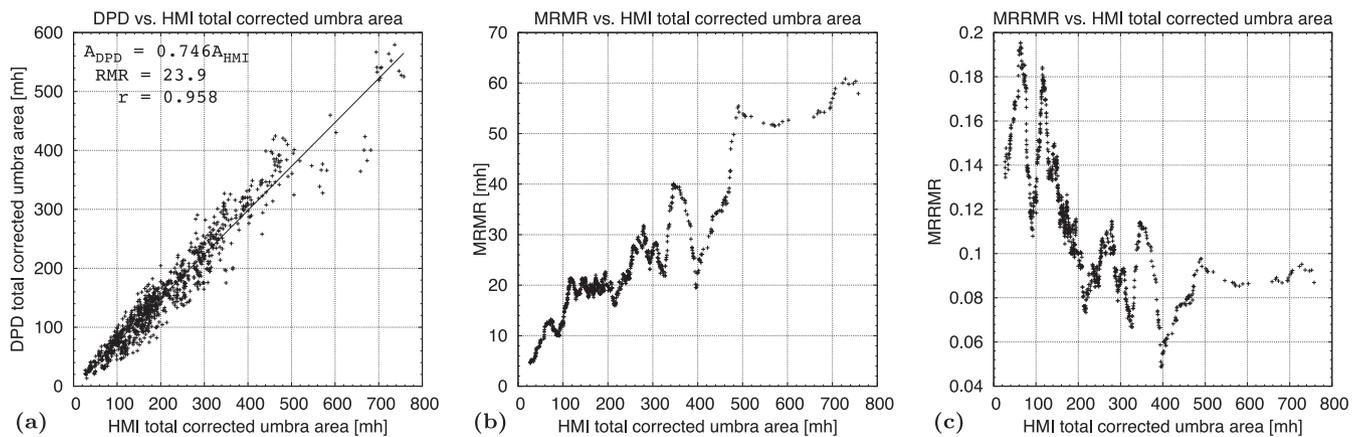

**Figure 6.** (a) Total corrected umbra areas in millionths of the solar hemisphere (*mh*) for 2012, for DPD versus HMI (pluses) and the regression line fitted to the data (solid line). (b) MRMR versus HMI total corrected umbra areas. (c) MRRMR versus HMI total corrected umbra areas. DPD images for the very good, the good and the medium seeing quality are used to create these figures and no selections are applied to the umbra area and to the heliocentric angle ($A > 0\,mh$ and $\gamma < 90°$).

**Table 7.** Summary of the results of the comparison of various sunspot areas measured in HMI and in DPD images. In this comparison, only DPD images with very good, good and medium seeing quality were used. The values in the table are the parameters characterizing the regression: $S$, slope of the regression lines; RMR, root mean square residual; $r$, correlation coefficient; $n$, sample size. The independent variable of the regression is the HMI area. The acronyms in the first column are: AT, area type; ICSA, individual corrected spot area; ICUA, individual corrected umbra area; TCSA, total corrected spot area; TCUA, total corrected umbra area.

| AT | $S$ | RMR [$mh$] | $r$ | $n$ |
|---|---|---|---|---|
| ICSA | 1.002 | 7.5 | 0.997 | 30 153 |
| ICUA | 0.928 | 2.1 | 0.987 | 14 403 |
| TCSA | 0.895 | 108.3 | 0.978 | 985 |
| TCUA | 0.746 | 23.9 | 0.958 | 985 |

**Table 8.** Statistics of the heliographic differences in latitude ($B$) and longitude ($L$) between identified pores and umbrae from HMI and from DPD (DPD-HMI). The sample size is 33 061.

| | $\Delta B$ | $\Delta L$ |
|---|---|---|
| Mean, $m$ [°] | 0.003 | −0.028 |
| Standard deviation, $\sigma$ [°] | 0.067 | 0.086 |

figure, we observe that it does not really characterize the accuracy of the regression. Its value is too low, especially for the individual area comparison. This is because the sample is biased towards small spots, i.e. there are many more smaller spots and umbrae than larger ones.

It is worth mentioning that the values (slopes) in the second column of Table 7 are consistent with those of the corresponding tables with the same area type in the sections above, if we perform averaging for the very good, the good and the medium seeing.

### 4.5 Comparison of the heliographic positions of the automated DPD and the HMI sunspots

Table 8 shows the statistics of the differences in the heliographic latitude ($B$) and longitude ($L$) between DPD and HMI (DPD-HMI) for the identified pores and umbrae. $m$ and $\sigma$ denote the mean and the standard deviation, respectively. The sample size is 33 061. The mean difference between the positions of the pores and umbrae in $L$ in the two catalogues is only $-0.028°$ and the mean latitude deviation is practically zero.

The $\sigma$ values are nearly the same (0.067°) for $\Delta B$ and (0.086°) for $\Delta L$. The high $\sigma$ values (relative to the means) can be explained by the fact that the positions of the DPD pores and small umbrae could be randomly changed by the seeing.

However, the slight mean heliographic coordinate difference, in itself, does not guarantee a slight alignment difference between the two sets of images, as we will see below.

The propagation of the alignment (orientation) error ($\Delta \tau$) of the solar image into the heliographic latitude ($B$) of a sunspot can be written as follows (Győri 2005):

$$\Delta B = \cos B_{\rm o} \sin L_{\rm cm} \Delta \tau, \quad (1)$$

where $B_{\rm o}$ is the heliographic latitude of the centre of the solar disc, $L_{\rm cm}$ is the central meridian distance of the sunspot, and $\Delta B$ is the error in $B$. Now, if we consider the $i$th identified spot pair, then from equation (1) we have

$$\Delta B_i^{\rm d} = \cos B_{\rm o,i}^{\rm d} \sin L_{\rm cm,i}^{\rm d} \Delta \tau_i^{\rm d}, \quad (2)$$

$$\Delta B_i^{\rm h} = \cos B_{\rm o,i}^{\rm h} \sin L_{\rm cm,i}^{\rm h} \Delta \tau_i^{\rm h}, \quad (3)$$

where the superscripts d and h stand for DPD and HMI images, respectively.

As the two images in the pair are close in time, so we may suppose that the differences between $B_{\rm o}^{\rm d}$ and $B_{\rm o}^{\rm h}$ as well as between $L_{\rm cm,i}^{\rm d}$ and $L_{\rm cm,i}^{\rm h}$ can be neglected. Additionally, we suppose that $\Delta \tau_i^{\rm d}$ and $\Delta \tau_i^{\rm h}$ do not depend on time, i.e. they are different constants ($\Delta \tau^{\rm d}$ and $\Delta \tau^{\rm h}$).

With these conditions in mind, subtracting equation (3) from equation (2), and summing for all the image pairs ($n$), we obtain after some algebra

$$\sum_{i=1}^{n} \Delta B_i = \Delta \tau \sum_{i=1}^{n} \cos B_{\rm o,i}^{\rm h} \sin L_{\rm cm,i}^{\rm h}, \quad (4)$$

where $\Delta B_i = \Delta B_i^{\rm d} - \Delta B_i^{\rm h} = B_i^{\rm d} - B_i^{\rm h}$ is the heliographic latitude difference of the spot for the $i$th image pair, and $\Delta \tau = \Delta \tau^{\rm d} - \Delta \tau^{\rm h}$ is the alignment difference between the DPD and the HMI images. From equation (4) we obtain

$$\Delta \tau = \sum_{i=1}^{n} \Delta B_i / \sum_{i=1}^{n} \cos B_{\rm o,i}^{\rm h} \sin L_{\rm cm,i}^{\rm h}. \quad (5)$$

As the right-hand side of equation (4) is antisymmetric in $L_{\rm cm}$, so significant cancellation of the terms in the right-hand side sum can occur. Therefore, the mean latitude deviation between the two catalogues can be small even when $\Delta \tau$ is high. To avoid this when determining $\Delta \tau$, we should separately determine $\Delta \tau$ for spots with $L_{\rm cm} > 0$ and with $L_{\rm cm} < 0$. By using equation (5), we separately determined $\Delta \tau$ for spots with $L_{\rm cm} > 0$ and with $L_{\rm cm} < 0$. Their average value is 0.019°. We accept this as the alignment difference between the DPD and the HMI images, i.e. $\Delta \tau = 0.019°$. The actual meaning of this angle is that the HMI images are rotated by 0.019° relative to the DPD images (or that the DPD images are rotated by $-0.019°$ relative to the HMI ones).

In an earlier investigation (Győri 2012), we found a 0.22° discrepancy (the MDI images are rotated by 0.22° relative to the HMI ones) between the alignments of the SOHO/MDI and the SDO/HMI images. Above, we have shown that the alignment difference between the DPD and SDO/HMI images is as small as 0.019°. This makes it more probable that the alignment discrepancy between the MDI and the HMI images can be attributed to the misalignment of the SOHO/MDI images.

As a counter check, we determined the rotation angle between the DPD and the MDI images for 2000, 2001 and 2002. We proceeded the same way as described above and obtained that the MDI images are rotated by 0.197° relative to the DPD ones. This result also corroborates the misalignment of the MDI images.

## 5 HMI AND REVISED DPD SUNSPOT AREA COMPARISON

As mentioned in Section 2, we strive to revise the DPD catalogues humanly too. When revising the DPD catalogue for 2012, we were in a pleasant situation of having near-simultaneous HMI images for comparing with the DPD images. This was especially useful when deciding about keeping or not keeping suspicious small pores and small umbrae in the DPD images. Therefore, the revised DPD (DPDr, in the following parts of the paper, the subscript r means revised) catalogue cannot be considered entirely independent of HMI, but this procedure allows us to test the quality of the revised DPD.

Table 9. First row: Comparison of the daily total corrected sunspot areas measured in HMI images and recorded in the revised DPD. The values in the table are the slopes of the regression lines. The independent variable is the HMI area. Second row: Averages computed for seeing in the range VG–M from Table 5. See the text for details.

| $A > 0\,mh$ $\gamma < 90°$ | $A > 5\,mh$ $\gamma < 90°$ | $A > 0\,mh$ $\gamma < 60°$ | $A > 5\,mh$ $\gamma < 60°$ |
|---|---|---|---|
| 0.950 | 0.965 | 0.983 | 1.005 |
| 0.895 | 0.933 | 0.949 | 0.999 |

Table 10. First row: Comparison of the daily total corrected umbra areas measured in HMI images and recorded in the revised DPD. The values in the table are the slopes of the regression lines. The independent variable is the HMI area. Second row: Averages computed for seeing in the range VG–M from Table 6. See the text for details.

| $A > 0\,mh$ $\gamma < 90°$ | $A > 5\,mh$ $\gamma < 90°$ | $A > 0\,mh$ $\gamma < 60°$ | $A > 5\,mh$ $\gamma < 60°$ |
|---|---|---|---|
| 0.906 | 0.960 | 0.925 | 1.003 |
| 0.762 | 0.837 | 0.810 | 0.897 |

It is worth mentioning that as we have many images to choose from for a given day (12–18 per day) at Gyula, thus, for the majority of the days of 2012, we succeeded in choosing images that fall into the categories of very good, good and medium seeing qualities.

The first row of Table 9 contains the slopes of the regression line for the DPDr versus the HMI daily total corrected sunspot areas with different kinds of selection applied to the DPDr and the HMI spots. If no selection is applied (column 1), then the DPDr daily total corrected sunspot areas are 5 per cent smaller than the HMI ones. If the spots with area smaller than 5 mh are left out (column 2), then the DPDr area deficit is 3.5 per cent. If the spots in the outer part of the solar disc are left out (column 3), then the area deficit is even smaller (only 1.7 per cent). Now if we leave out the spots with area smaller than 5 mh and also the ones in the outer part of the solar disc (column 4), then the two areas are practically the same.

In the first row of Table 10, we see the slopes of the regression lines for the DPDr and the HMI daily total corrected sunspot umbra areas with various selections applied to the umbrae. If no selection is applied (column 1), then the DPDr daily total corrected umbra areas are 9 per cent smaller than the HMI ones. If the umbrae with area smaller than 5 mh are left out (column 2), then the DPDr area deficit is 4 per cent. If the spots in the outer part of the solar disc are left out (column 3), then the area deficit is 7.5 per cent. Now if we leave out the umbrae with area smaller than 5 mh and also the ones in the outer part of the solar disc (column 4), then the two areas are practically the same.

We provide a second row in Tables 9 and 10 to show how the revision of the DPD improves the agreement between DPD and HMI total sunspot and total umbra areas. These second rows, therefore, contain the averages of the slopes for the total sunspot area and for the total umbra area computed for seeing in the range VG–M from Tables 5 and 6, respectively. A comparison of the corresponding values in the first and second rows shows that the revision significantly improved the agreement between the two catalogues. In practice, when no selection is applied, the difference between the automated DPD and HMI is divided by 2 when using the revised version.

We would like to emphasize that this significant improvement can be attributed to the near-simultaneous superior-quality HMI images used during the revision of the DPD as controlling images. This is an exceptional situation. When a revision is performed without this possibility, we cannot expect such significant improvement.

Fig. 7 shows the details of the regression between total corrected spot areas from the revised DPD catalogue and from nearby HMI images. Fig. 7(a) shows the scatter plot and the regression line fitted to the data. Fig. 7(b) shows MRMR versus HMI total corrected spot areas. Fig. 7(c) shows MRRMR versus HMI total corrected spot areas. Fig. 8 shows the same plots for the total corrected umbra areas.

If we compare Fig. 5 with Fig. 7 and then Fig. 6 with Fig. 8, we see how the selection of the best observation and the revision of the spot data improve the fit (decrease the scatter around the regression line) of DPD total sunspot and total umbra areas to those of HMI, and how they decrease the values of MRMR and MRRMR.

## 6 COMPARISON OF THE SUNSPOT AREAS FROM ROYAL GREENWICH OBSERVATORY AND DHO IMAGES FOR 1974, 1975 AND 1976

In this section, we deal with three catalogues: the DPD based on Debrecen images, the Greenwich Photoheliographic Data (GPD) based on digitized Royal Greenwich Observatory (RGO) images and compiled with the same routines as the DPD, and the GPR[1] based on the same images as the GPD and published by Greenwich up to 1976. To compare the three catalogues, we produced a revised version of the DPD and GPD catalogues for 1974, 1975 and 1976.

As the revision revealed for these years, some mistakes were made during the digitalization process of the RGO images, and so about 10 per cent of the digitized images were erroneous in some way. There were mistakes in the time of the observation (included in the image file name). Moreover, there were images with a bad alignment of the film with the CCD camera, e.g. the solar east is on the right side of the image instead of the intended west side.

These errors can be revealed by checking the time variation of the heliographic positions of the spots. Higher leaps in the positions of the spots indicate errors in the time of the observation or in the alignment of the photoheliogram and the CCD camera. After finding an error, we need to decide the type of error. An alignment error can be found by visually comparing the images on consecutive days. If the cause of the error is a bad alignment, the image is rotated or reflected properly by the processing software (SAM). For a time error, we proceed as follows.

The effect of an erroneous observation time for a solar image mainly appears in the *L* (heliographic longitude) coordinate of the sunspots. For example, an error of 1 h causes 0.6° error in *L* ($B = 16°$, the error has some dependence on *B*). By using two instances of SAMm, the GPD heliographic coordinates of the sunspots were checked against those of DPD. By clicking with the mouse on a sunspot, SAMm writes out a lot of data about the spot, among them is *L*. The user of SAMm looked for a systematic difference in *L* for those spots she managed to identify with each other in the image pair. If the user of SAMm found such a difference, she changed the observation time of the GPD image and ran SAM again on this image. The procedure was repeated until no systematic difference in *L* was found.

---

[1] Royal Greenwich Observatory, 1874–1976, Greenwich Photoheliographic Results [Greenwich Observations (1874–1955); Royal Greenwich Observatory Bulletins (1956–1961); Royal Observatory Annals (1962–1976)], in 91 volumes, Edinburgh and Eastbourne. Available at: http://www.ngdc.noaa.gov/stp/solar/sunspotregionsdata.html

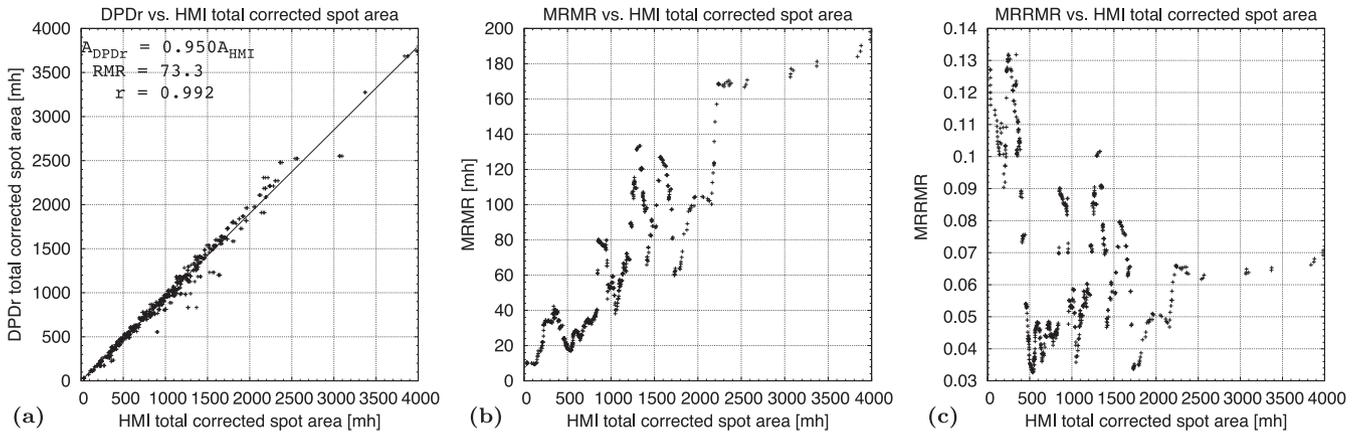

**Figure 7.** (a) Total corrected spot areas in millionths of the solar hemisphere (*mh*) for 2012, for DPDr versus HMI (pluses) and the regression line fitted to the data (solid line). (b) MRMR versus HMI total corrected spot areas. (c) MRRMR versus HMI total corrected spot areas. No selections are applied for the spot area and for the heliocentric angle ($A > 0$ *mh* and $\gamma < 90°$).

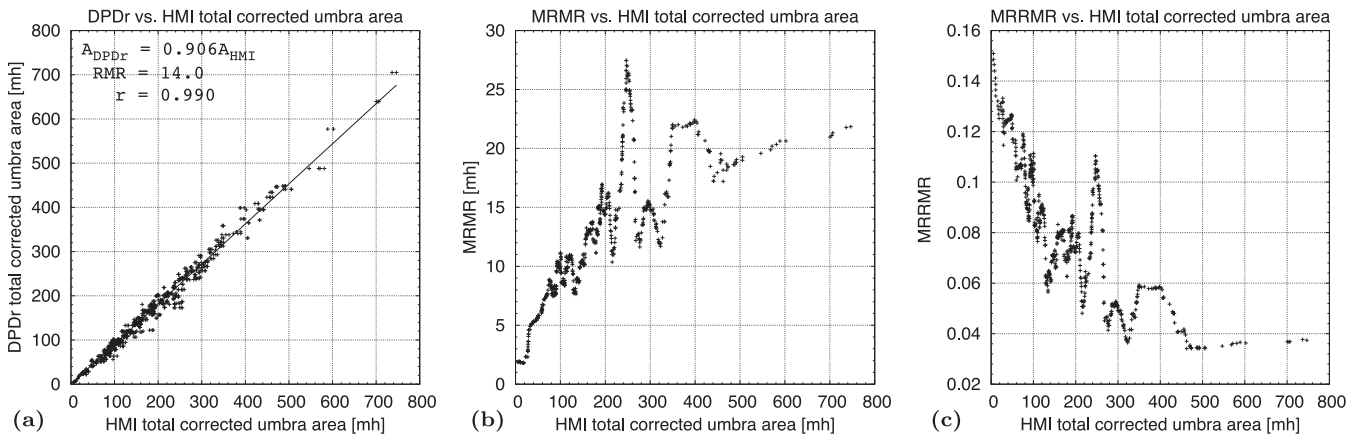

**Figure 8.** (a) Total corrected umbra areas in millionths of the solar hemisphere (*mh*) for 2012, for DPDr versus HMI (pluses) and the regression line fitted to the data (solid line). (b) MRMR versus HMI total corrected umbra areas. (c) MRRMR versus HMI total corrected umbra areas. No selections are applied for the umbra area and for the heliocentric angle ($A > 0$ *mh* and $\gamma < 90°$).

The accuracy of this procedure depends on the number of identified sunspots in the two images, and on the true time difference (proper motion of the sunspots) between the two images. We estimate that its time accuracy is in the range 15–60 min and its corresponding accuracy in *L* is in the range 0.15–0.6°.

The formula that transforms one pixel area in the solar image on to the solar surface contains three quantities: the distance of the pixel from the centre of the solar disc, the radius of the solar disc and the semi-view angle of the Sun. From these quantities, only the semi-view angle depends on the observation time. The other two quantities are measured in the solar image itself. However, the semi-view angle is practically constant for any given day. Therefore, we may say that the measured sunspot area is not influenced by an error in the observation time.

The above argument is valid if the Sun is not near the horizon according to the erroneous observation time. If this is the case, then a false differential refraction correction is performed on the solar disc. This distorts the solar disc. However, on examining the shape of the solar disc after the differential refraction correction (it should be a circle), SAM notices this problem and signals it to the user. If this is the case, then the user successively changes the observation time until SAM accepts the differential refraction correction. For how SAM performs the differential refraction correction, see Győri (1993).

There are several differences between RGO and DHO images. The RGO photographic images are 19 cm in diameter, while the DPD ones are only 10 cm in diameter. The emulsion type and its gamma are probably different between the two sets of images. For the GPD catalogue, the RGO photoheliograms were digitized by a CCD camera leading to digitized images with an image scale of ∼0.57 arcsec pixel$^{-1}$, while the DPD images were digitized by a scanner, resulting in digitized images with an image scale of ∼0.26 arcsec pixel$^{-1}$. The analysis to produce the two catalogues (DPD and GPD) was, however, similar (based on SAM), unlike the old procedure to produce the GPR catalogue.

The average seeing for the images used to produce DPD is much better than the average seeing for the images used in GPR and GPD. This has two reasons. First, several photoheliograms were taken each day in Gyula, allowing us to select the best one each day. Second, the Gyula heliograph is situated in a grassy park on the top of a water tower about 40 m above the ground, so that at this height the refractive index fluctuation induced by the uneven heating of the ground by the Sun is significantly damped, yielding better seeing conditions.

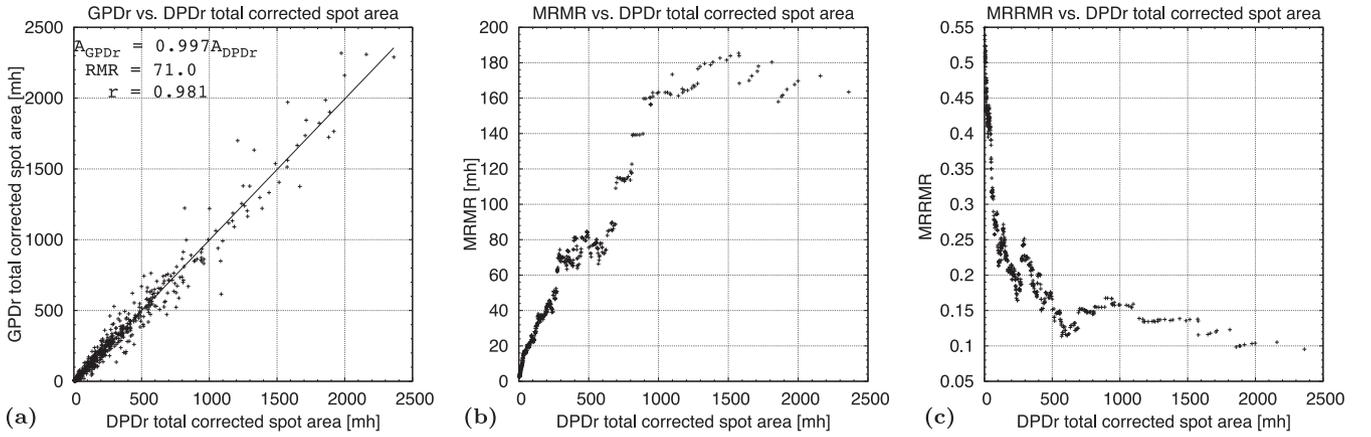

**Figure 9.** (a) Daily corrected total sunspot areas in millionths of the solar hemisphere (*mh*) for 1974, 1975 and 1976 for GPDr versus DPDr (pluses) and the regression line fitted to the data (solid line). (b) MRMR versus DPDr total corrected spot areas. (c) MRRMR versus DPDr total corrected spot areas.

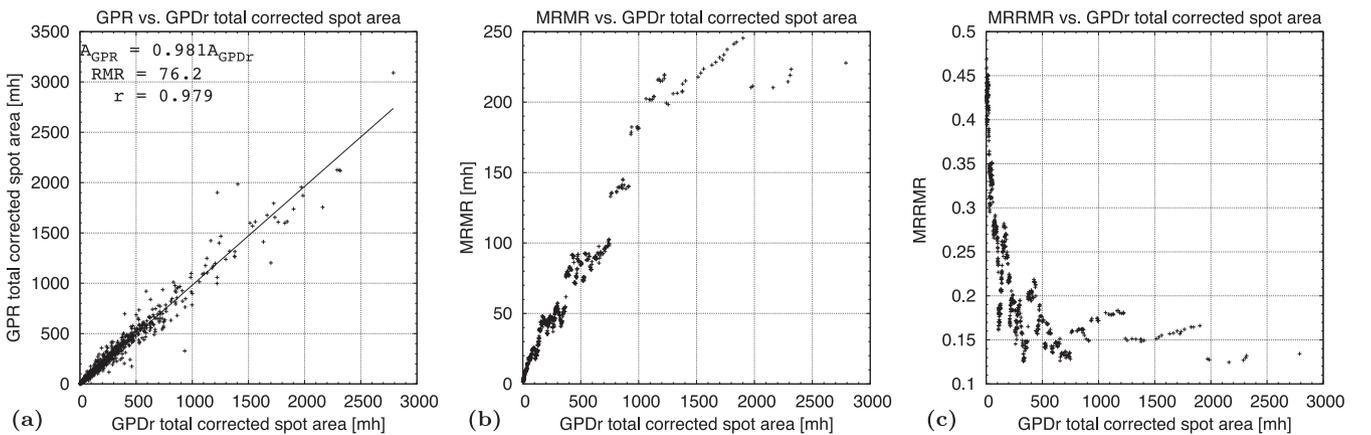

**Figure 10.** (a) Daily corrected total sunspot areas in millionths of the solar hemisphere (*mh*) for 1974, 1975 and 1976 for GPR versus GPDr (pluses) and the regression line fitted to the data (solid line). (b) MRMR versus GPDr total corrected spot areas. (c) MRRMR versus GPDr total corrected spot areas.

On average, the images used to produce DPD are well exposed. In contrast, on average, the images used to produce GPR and GPD are overexposed.

The differences in the image quality between DPD and GPD show up in the various feature numbers. There are 3.3 times more pores, 1.7 times more penumbrae (spots with a penumbral structure and with one or more umbrae), and 2.25 times more umbrae in the DPD than in the GPD.

In the following sections, we examine whether these differences have any effect on the measured area of sunspots.

### 6.1 Revised GPD and revised DPD daily total corrected sunspot area comparison

Fig. 9 shows the details of the regression between 583 daily total corrected spot areas from the GPDr ($A_{\text{GPDr}}^{\text{tcs}}$) and from the DPDr ($A_{\text{DPDr}}^{\text{tcs}}$) catalogues. Fig. 9(a) shows the scatter plot and the regression line fitted to the data. Fig. 9(b) shows MRMR versus DPDr total corrected spot areas. Fig. 9(c) shows MRRMR versus DPDr total corrected spot areas.

The equation, the root mean square residual (RMR) of the regression line and the correlation coefficient (*r*) are:

$$A_{\text{GPDr}}^{\text{tcs}} = (0.997 \pm 0.006) A_{\text{DPDr}}^{\text{tcs}},$$
$$\text{RMR} = 71.0,$$
$$r = 0.981. \quad (6)$$

As equation (6) shows, there is, therefore, no systematic difference between the GPDr and the DPDr daily total corrected sunspot areas. This is an unexpected result because, as we have seen before, there are significant differences in the average seeing, in the average exposure and in the feature numbers between the GPDr and DPDr. It seems that the area decrease caused by overexposure and by the lower pore and penumbra number are compensated for by the area increase caused by the poorer seeing for GPDr.

### 6.2 GPR and revised GPD daily total corrected sunspot area comparison

The Greenwich sunspot catalogue has been widely used in the literature and, therefore, it is useful to compare it with the newer version we propose here (GPDr).

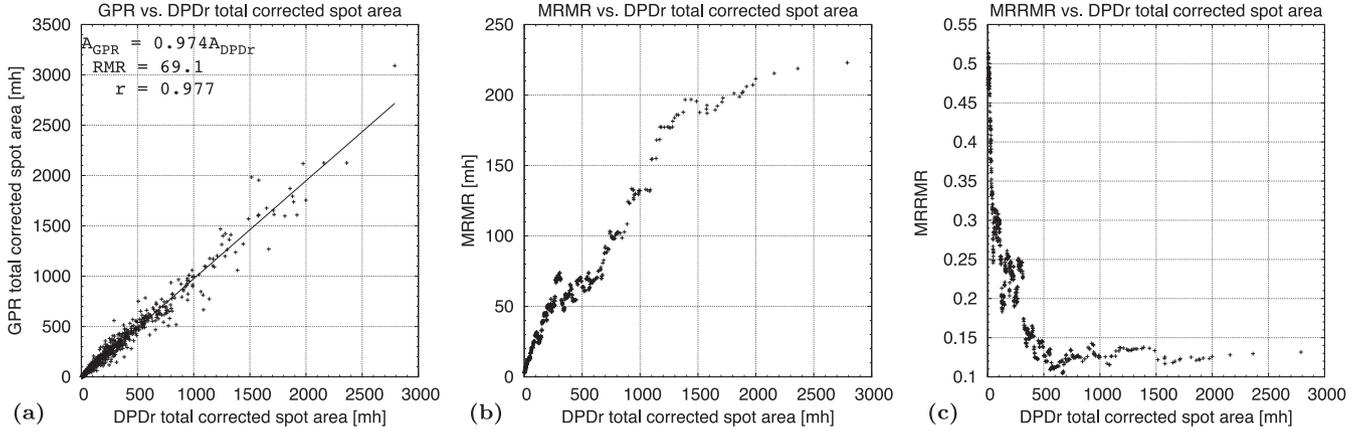

**Figure 11.** (a) Daily corrected total sunspot areas in millionths of the solar hemisphere (*mh*) for 1974, 1975 and 1976 for GPR versus DPDr (pluses) and the regression line fitted to the data (solid line). (b) MRMR versus DPDr total corrected spot areas. (c) MRRMR versus DPDr total corrected spot areas.

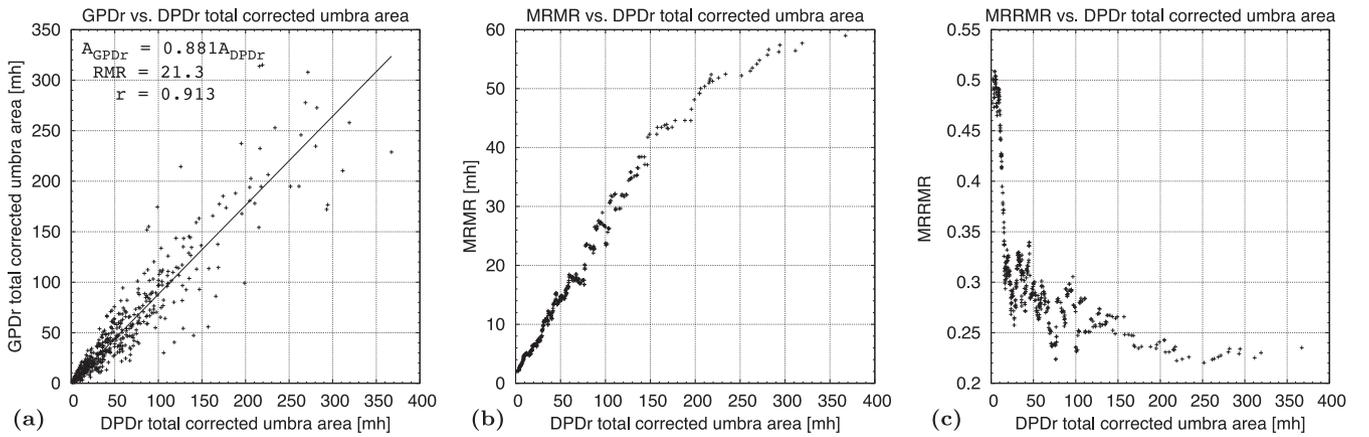

**Figure 12.** (a) Daily corrected total umbra areas in millionths of the solar hemisphere (*mh*) for 1974, 1975 and 1976 for GPDr versus DPDr (pluses) and the regression line fitted to the data (solid line). (b) MRMR versus DPDr total corrected umbra areas. (c) MRRMR versus DPDr total corrected umbra areas.

Fig. 10 shows the details of the regression between 667 daily total corrected spot areas from the GPR ($A_{\rm GPR}^{\rm tcs}$) and from the GPDr ($A_{\rm GPDr}^{\rm tcs}$) catalogues. Fig. 10(a) shows the scatter plot and the regression line fitted to the data. Fig. 10(b) shows MRMR versus GPDr total corrected spot areas. Fig. 10(c) shows MRRMR versus GPDr total corrected spot areas. The equation, RMR of the regression line and the correlation coefficient ($r$) are:

$A_{\rm GPR}^{\rm tcs} = (0.981 \pm 0.006) A_{\rm GPDr}^{\rm tcs},$
RMR = 76.2,
　$r = 0.979.$ (7)

The GPR daily total corrected sunspot areas are about 2 per cent smaller than the GPDr ones. If we exclude spots with area smaller than 5 *mh* from the GPDr catalogue when determining the total spot areas, then we have:

$A_{\rm GPR}^{\rm tcs} = (0.998 \pm 0.006) A_{\rm GPDr}^{\rm tcs},$
RMR = 80.4,
　$r = 0.977.$ (8)

In this case, the GPR and GPDr areas are statistically the same. This result also indicates that, in GPR, the smaller spots were overlooked from some considerations compared to the re-analysis of the same images we have performed, notwithstanding that they can be observed in the solar images.

### 6.3 GPR and revised DPD daily total corrected sunspot area comparison

Fig. 11 shows the details of the regression between daily total corrected spot areas from the GPR ($A_{\rm GPR}^{\rm tcs}$) and from the DPDr ($A_{\rm DPDr}^{\rm tcs}$) catalogues. Fig. 11(a) shows the scatter plot and the regression line fitted to the data. Fig. 11(b) shows MRMR versus GPDr total corrected spot areas. Fig. 11(c) shows MRRMR versus GPDr total corrected spot areas.

The equation, RMR of the regression line and the correlation coefficient ($r$) are:

$A_{\rm GPR}^{\rm tcs} = (0.974 \pm 0.005) A_{\rm DPDr}^{\rm tcs},$
RMR = 69.1,
　$r = 0.977.$ (9)

The sample size is 673. The GPR daily total corrected sunspot areas are about 2.5 per cent smaller than those of the DPDr. Notice that this result is consistent (as expected) with the results of the previous two sections.

### 6.4 Revised GPD and revised DPD daily total corrected umbra area comparison

Fig. 12 depicts the details of the regression between daily total corrected umbra areas from the GPDr ($A_{\rm GPDr}^{\rm tcu}$) and from the DPDr ($A_{\text{DPDr}}^{\text{tcu}}$) catalogues. Fig. 12(a) shows the scatter plot and the regression line fitted to the data. Fig. 12(b) shows MRMR versus DPDr total corrected spot areas. Fig. 12(c) shows MRRMR versus DPDr total corrected spot areas.

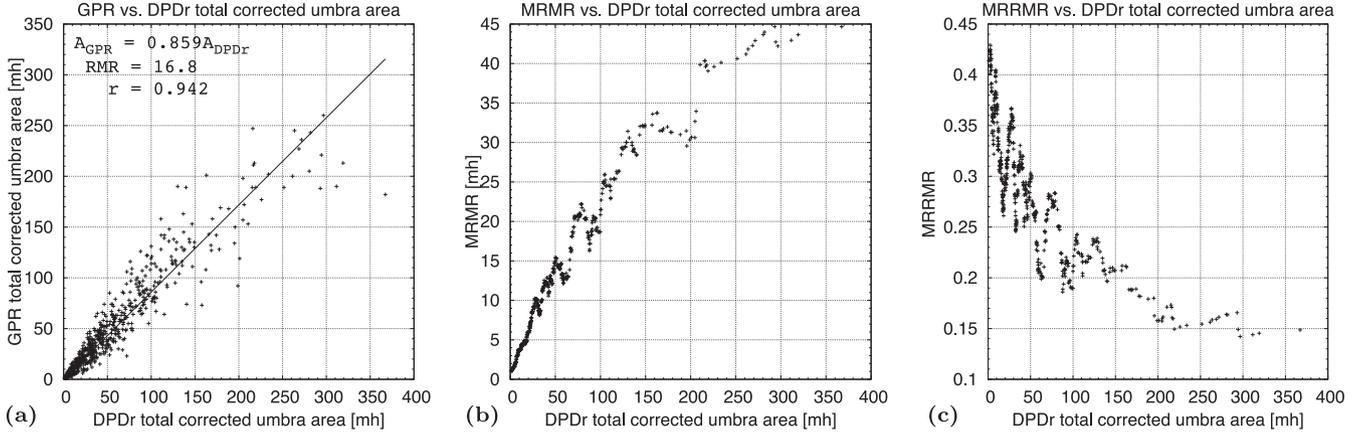

**Figure 13.** (a) Daily total corrected umbra areas in millionths of the solar hemisphere (*mh*) for 1974, 1975 and 1976 for GPR versus DPDr (pluses) and the regression line fitted to the data (solid line). (b) MRMR versus DPDr daily total corrected umbra areas. (c) MRRMR versus DPDr daily total corrected umbra areas.

The equation, RMR of the regression line and the correlation coefficient (*r*) are:

$$A_{\text{GPDr}}^{\text{tcu}} = (0.881 \pm 0.011) A_{\text{DPDr}}^{\text{tcu}},$$
$$\text{RMR} = 21.3,$$
$$r = 0.913. \quad (10)$$

The sample size is 477. The GPDr daily total corrected umbra areas are about 12 per cent smaller than the DPDr ones. If we leave out umbrae with an area smaller than 5 *mh* for both catalogues when determining the total umbra areas, then we have:

$$A_{\text{GPDr}}^{\text{tcu}} = (0.985 \pm 0.016) A_{\text{DPDr}}^{\text{tcu}},$$
$$\text{RMR} = 22.4,$$
$$r = 0.866. \quad (11)$$

In this case, we experience that the GPDr daily total corrected umbra areas are just 1.5 per cent smaller than the DPDr ones. This indicates that the smaller umbrae do not separate from the penumbra in the GPDr. The cause may be the lower scale of the GPDr images and that the average seeing condition is poorer for GPDr than for DPDr.

### 6.5 GPR and revised DPD daily total corrected umbra area comparison

Fig. 13 depicts the details of the regression between 617 daily total corrected umbra areas from the GPDr ($A_{\text{GPDr}}^{\text{tcu}}$) and from the DPDr ($A_{\text{DPDr}}^{\text{tcu}}$) catalogues. Fig. 12(a) shows the scatter plot and the regression line fitted to the data. Fig. 12(b) shows MRMR versus DPDr daily total corrected umbra areas. Fig. 12(c) shows MRRMR versus DPDr daily total corrected umbra areas.

The equation, RMR of the regression line and the correlation coefficient (*r*) are

$$A_{\text{GPR}}^{\text{tcu}} = (0.859 \pm 0.011) A_{\text{DPDr}}^{\text{tcu}},$$
$$\text{RMR} = 16.8,$$
$$r = 0.942. \quad (12)$$

The GPR daily total corrected umbra areas are about 14 per cent smaller than the DPDr ones.

**Table 11.** Statistics of the differences in heliographic latitude (*B*) and longitude (*L*) between identified pores and umbrae from GPD and from DPD (GPD-DPD).

|  | 1974 | | 1975 | | 1976 | |
| --- | --- | --- | --- | --- | --- | --- |
|  | Δ*B* | Δ*L* | Δ*B* | Δ*L* | Δ*B* | Δ*L* |
| *m* [°] | 0.04 | −0.05 | 0.01 | −0.04 | −0.01 | −0.04 |
| *σ* [°] | 0.21 | 0.23 | 0.20 | 0.21 | 0.21 | 0.21 |
| *n* | 2992 | 2992 | 1152 | 1152 | 955 | 955 |

*m*: mean, *σ*: standard deviation, *n*: sample size.

### 6.6 Comparison of the heliographic positions of revised GPD and revised DPD sunspots

Table 11 shows the statistics of the differences in the heliographic latitude (*B*) and longitude (*L*) between GPD and DPD (GPD-DPD) for the identified pores and umbrae. *m*, *σ* and *n* denote the mean, the standard deviation and the sample size, respectively. We see that the systematic difference between the positions of the pores and umbrae in the two catalogues is small, and it does not exceed 0.05° in absolute value.

The *σ* values are nearly the same (about 0.21°) for Δ*B* and Δ*L* for all the 3 yr. Part of the dispersion may be because the GPD and DPD images are not co-temporal. Generally, there are several hours difference in their observation times, and therefore some proper motions of the spots may occur. The positions of the pores and small umbrae could be randomly changed even by the seeing.

### 6.7 Testing Foukal's explanation for the large spot area difference in the Solar Observing Optical Network and GPR catalogues

There are large sunspot area differences (about 40 per cent) between NOAA/USAF Solar Observing Optical Network (SOON)[2] and GPR daily total sunspot area records (Baranyi et al. 2001, 2013; Hathaway, Wilson & Reichmann 2002). Foukal (2014) suggested that this can be can be explained by the SOON area measurement practice. Namely, that the small spots (<10 *mh*) were not directly measured but an area of 2 *mh* was assigned to each.

---

[2] NOAA NCEI National Geophysical Data Center Solar–Terrestrial Physics Division, Boulder, USAF network (USAF-MWL) data, 1981–2013. Available at: http://www.ngdc.noaa.gov/stp/solar/sunspotregionsdata.html

This explanation, as noted by Foukal, cannot be tested against GPR catalogues because they do not contain the areas of the individual small spots. However, as GPDr provides the areas of the individual spots as well, we can simulate the effect by modifying the GPDr spot catalogue.

Thus, from the GPDr catalogue, we produced two more catalogues: GPDr$_2$ and GPDr$_0$. To derive GPDr$_2$ and GPDr$_0$ from GPDr, the area of the spots in GPDr with area smaller than 10 $mh$ were replaced by 2 $mh$ and by 0 $mh$, respectively. equation (13) summarizes the regression between GPDr and GPDr$_2$ daily total corrected spot areas:

$$A_{\mathrm{GPDr}_2}^{\mathrm{tcs}} = (0.985 \pm 0.000) A_{\mathrm{GPDr}}^{\mathrm{tcs}},$$
$$\mathrm{RMR} = 7.9,$$
$$r = 1.000. \qquad (13)$$

From equation (13), we see that the daily total sunspot areas recorded by GPDr$_2$ are 1.5 per cent smaller than those recorded by GPDr. This is far away from the expected 40 per cent.

Now let us see what happens when we leave out all the spots with areas smaller than 10 $mh$ from GPDr, i.e. when we compare GPDr and GPDr$_0$:

$$A_{\mathrm{GPDr}_0}^{\mathrm{tcs}} = (0.966 \pm 0.001) A_{\mathrm{GPDr}}^{\mathrm{tcs}},$$
$$\mathrm{RMR} = 17.6,$$
$$r = 0.999. \qquad (14)$$

In this case, as equation (14) exhibits, GPDr$_0$ daily total sunspot areas are 3.4 per cent smaller than GPDr ones. Even this result is far from the 40 per cent.

Moreover, it is worth recalling here that GPR daily total corrected spot areas are 1.6 per cent smaller than GPDr ones, and that this area deficit can be explained by RGO measurers overlooking small spots (see Section 6.2).

We conclude that the 40 per cent is due to the different treatment of large structures (for example, to a completely different way of determining spot borders).

## 7 SETTING UP A LONG-TERM SUNSPOT DATA BASE FROM GPR, REVISED DPD, AND HMI CATALOGUES

If we want to draw up a long-term homogeneous sunspot area data base from the GPR, DPD and HMI catalogues, we need to know how to convert total sunspot and umbra areas between these catalogues. The quality of the HMI images is better than those of GPR and DPD, so it is reasonable to convert GPR and DPD areas to HMI ones. Now, using the results of the previous sections, we can determine the necessary conversion factors.

### 7.1 From DPD to HMI

Taking the reciprocal of the element in the third row and the second column of Table 7, we get the conversion factor for converting automated DPD (DPDa) total corrected sunspot areas into those of HMI. Using this factor, the conversion equation can be written as

$$A_{\mathrm{HMI,a}}^{\mathrm{tcs}} = 1.12 A_{\mathrm{DPDa}}^{\mathrm{tcs}}. \qquad (15)$$

Taking the reciprocal of the element in the fourth row and the second column of Table 7, we get the conversion factor for converting DPDa total corrected umbra areas into those of HMI. Using this factor, the conversion equation can be written as

$$A_{\mathrm{HMI,a}}^{\mathrm{tcu}} = 1.34 A_{\mathrm{DPDa}}^{\mathrm{tcu}}. \qquad (16)$$

Taking the reciprocal of the element in the first row and the first column of Table 9, we get the conversion factor for converting DPDr total corrected sunspot areas into those of HMI. Using this factor, the conversion equation can be written as

$$A_{\mathrm{HMI,r}}^{\mathrm{tcs}} = 1.05 A_{\mathrm{DPDr}}^{\mathrm{tcs}}. \qquad (17)$$

Taking the reciprocal of the element in the first row and the first column of Table 10, we get the conversion factor for converting DPDr total corrected umbra areas into those of HMI. Using this factor, the conversion equation can be written as

$$A_{\mathrm{HMI,r}}^{\mathrm{tcu}} = 1.10 A_{\mathrm{DPDr}}^{\mathrm{tcu}}. \qquad (18)$$

We mentioned earlier that the good quality, seeing-free and practically co-temporal HMI images provided an exceptionally good possibility for revising the DPD catalogue for 2012. However, since we do not have such good solar images that cope with the HMI ones for controlling purposes for the years before 2012, therefore the revision of the DPD for these years may not result in such an extensive improvement. In other words, the revised DPD has been tested on only 2012, so it may not be as good all the time.

Based on this, we think that the conversion factors between the DPDr and the HMI total corrected sunspot and umbra areas for 2012 underestimate the conversion factors for other years. In our experience, the DPDa total corrected sunspot and umbra areas are statistically smaller than the DPDr ones for the years other than 2012, too. Based on these two statements, we guess that the conversion factors between the DPDr and the HMI total corrected sunspot and umbra areas, for years other than 2012, are somewhere between the DPDa and the DPDr conversion factors for 2012. Under the given circumstances, we believe that the most we can do is to accept the mean of the automatic and the revised conversion factors (equations 15 and 17 for the total corrected spot area, and equations 16 and 18 for the total corrected umbra area) as the conversion factor between the HMI and the revised DPD areas for the years other than 2012. That is, we can write

$$A_{\mathrm{HMI}}^{\mathrm{tcs}} = 1.08 A_{\mathrm{DPDr}}^{\mathrm{tcs}} \qquad (19)$$

for the total corrected sunspot area, and

$$A_{\mathrm{HMI}}^{\mathrm{tcu}} = 1.22 A_{\mathrm{DPDr}}^{\mathrm{tcu}} \qquad (20)$$

for the total corrected umbra area.

### 7.2 From GPR to HMI

Now, if we want to convert GPR area to HMI ones, we first need to convert GPR areas to DPDr ones. Inverting equation (9), we get the factor for converting GPR total corrected spot areas into those of DPDr. Using this factor, the conversion equation can be written as

$$A_{\mathrm{DPDr}}^{\mathrm{tcs}} = 1.03 A_{\mathrm{GPR}}^{\mathrm{tcs}}. \qquad (21)$$

Inverting equation (12), we get the factor for converting GPR total corrected umbra areas to those of DPD. Using this factor, the conversion equation can be written as

$$A_{\mathrm{DPDr}}^{\mathrm{tcu}} = 1.16 A_{\mathrm{GPR}}^{\mathrm{tcu}}. \qquad (22)$$

After having converted the GPR total corrected spot areas to the DPDr ones, we convert it to HMI ones using equation (19). The conversion equation is

$$A_{\mathrm{HMI}}^{\mathrm{tcs}} = 1.11 A_{\mathrm{GPR}}^{\mathrm{tcs}}. \qquad (23)$$

Similarly, after having converted the GPR total corrected umbra areas to the DPDr ones, we convert it to HMI ones using equation (20). The conversion equation is

$$A_{\text{HMI}}^{\text{tcu}} = 1.41 A_{\text{GPR}}^{\text{tcu}}. \quad (24)$$

## 8 DISCUSSION AND CONCLUSIONS

The measured areas of sunspots are influenced by several factors that determine the image quality. In this paper, we tried to unfold the effect of some of them, namely the seeing quality and the foreshortening of the solar image, on the measured spot and umbra areas (Section 4). For this purpose, we used the areas determined from the space-based (HMI) and ground-based (DPD) images over 2012.

If we compare the results (column 2 of Table 5 and column 2 of Table 6) obtained when comparing automatic and revised DPD with HMI, then we see that the revised DPD total corrected sunspot and umbra area deficits (relative to HMI) are smaller (about half) than those of the automatic DPD.

We found that the MDI images are rotated by 0.197° relative to the DPD ones. We also found a slight rotation between DPD and HMI images. DPD images are rotated by −0.019° relative to the HMI ones. In a previous paper (Győri 2012), we found that MDI images are rotated by 0.22° relative to the HMI ones. These findings suggest that the MDI images are misaligned by 0.22°.

We compared daily sunspot areas and positions of three catalogues (GPR, GPD and DPDr) for 1974, 1975 and 1976, and found good agreement among them (Section 6).

We showed (Section 6.7) that only a very small part (only 1.5 per cent instead of 40 per cent) of the SOON spot area deficit can be explained by the SOON area measurement practice, namely, that the spots smaller than 10 $mh$ were not directly measured but an area of 2 $mh$ was assigned to each.

We determined the conversion factors for converting DPDr and GPR total corrected spot areas and total corrected umbra areas into those of HMI. Table 12 summarizes these factors, which are useful for studying long-term changes (embracing 140 yr) in sunspot areas.

We underline here again that the conversion factors between the DPDr and the HMI total corrected sunspot and umbra areas and between the GPR and the HMI total corrected sunspot and umbra areas are speculative (except for 2012) in the sense that there is no direct evidence for them, so they should be used with caution (see Section 7.1). Such evidence would be obtained if we revised the DPD catalogue for 2012 without using the HMI images as control ones. However, that is time and human resource consuming, and, for the time being, we have no resources for this.

**Table 12.** Factors for converting DPDr and GPR total corrected spot (TCS) areas and total corrected umbra (TCU) areas into those of HMI.

|  | DPDr | | GPR | |
| --- | --- | --- | --- | --- |
|  | TCS | TCU | TCS | TCU |
| HMI | 1.08 | 1.22 | 1.11 | 1.41 |

## ACKNOWLEDGEMENTS

The SDO/HMI images are available by courtesy of NASA/SDO and the AIA, EVE and HMI science teams. SDO images are courtesy of the Solar Dynamic Observatory (NASA). RGO digitized images are available by courtesy of MSSL. This work was supported by the European Community's Seventh Framework Program (FP7 SP1-Cooperation) under grant agreement 284461 (EHEROES), by ESA PECS project C98081, and by the grant TÁMOP 4.2.2.C-11/1/KONV/2012-0015. We thank the referee for several suggestions, which made the paper clearer.

This paper has been typeset from a T<sub>E</sub>X/L<sup>A</sup>T<sub>E</sub>X file prepared by the author.